\documentclass[journal]{IEEEtran}
\usepackage[pdftex]{graphicx}
\usepackage{xcolor}
\begin{document}
%
\title{Integrated Photonic Functions Using Anisotropic 2D Material Structures}
\author{Po-Han Chang, Charles Lin, and Amr S. Helmy,~\IEEEmembership{Senior, Member,~IEEE,}

\thanks{P. Chang, C. Lin and A. Helmy are with the Edward S. Rogers Sr. of Department of Electrical and Computer Engineering, University of Toronto, Ontario, Canada.}
\thanks{Manuscript received April 19, 2005; revised August 26, 2015.}}

\markboth{Journal of \LaTeX\ Class Files,~Vol.~14, No.~8, August~2015}%
{Shell \MakeLowercase{\textit{et al.}}: Bare Demo of IEEEtran.cls for IEEE Journals}

\maketitle

\begin{abstract}
Plasmonic waveguides based on 2D materials, which enable the formations of guided modes confined around few-layered material, are promising plasmonic platforms for the miniaturization of photonic devices. Nonetheless, such waveguides support modes that are evanescent in the waveguide core with the majority of the fields concentrated around waveguide edges, which are different from those supported by 3D dielectric waveguides where the modal fields are of oscillatory nature and peak at the center. As a result, many photonic devices and functionalities that can be achieved within 3D dielectric waveguides based on total-internal-reflation modes cannot be realized using 2D material-based plasmonic structures. In this work, we propose and demonstrate how to leverage anisotropy in 2D materials to tailor of modal fields supported by 2D material waveguide for the first time. By regulating material absorption of the constituent 2D materials, the modal fields of these 2D modes can be tailored to localize around the waveguide center, which in turn can improve the efficiencies of coupling-based photonic functions using 2D materials, from in-plane multimode-interference couplers to out-of-plane optical radiation. Using natural anisotropic 2D materials such as black phosphorus, these pivotal functions can expand existing device capabilities that are typically achieved in 3D dielectrics but using 2D materials, thus allowing for the implementation of 2D plasmonic circuits with no need to relying on 3D layers.

\end{abstract}

\begin{IEEEkeywords}
2D materials, plasmonics, waveguides.
\end{IEEEkeywords}

\IEEEpeerreviewmaketitle

\section{Introduction}
Intensive research has been dedicated to utilize 2D materials for the realization of plasmonic waveguides \cite{Yu:2017,Garcia:2016,Alu:2016}. These waveguides enable the construction of components on the nanometer-scale with a superior degree of confinement. Most notably, the reduced dimensionality can provide enhanced light-matter-interaction and gate-tunable reconfigurability not attainable in 3D plasmonics, where bulk metal layers are utilized  \cite{Vakil:2011,Koppens:2011}. The confinement offered by such guides, can empower a plethora of functions ranging from giant nonlinearities at reduced power levels, to new levels of sensitivity in transduction elements \cite{Yu:2011,Christensen:2012,Farmer:2016,Sipe:2014}. The utilization of isotropic 2D materials, such as graphene, to realize 2D plasmonic waveguides with no supporting structures, dielectric or otherwise, has received the most attention, when compared to the use of its anisotropic counterparts \cite{Vakil:2011}. For isotropic 2D plasmonics of finite waveguide width; namely a 2D ribbon, the transverse modes are confined based on evanescent coupling between the two edge modes on either side of the film. As a result, the majority of the modal fields of the fundamental ribbon modes supported by isotropic 2D plasmonics are inherently concentrated near the edges of the film with the field in the core being evanescent. This is in contrast to modes in conventional 3D dielectric waveguides where the field in the core is oscillatory in nature and peaks in the center of the core. The guided modes in such cases are based on total internal reflection (TIR). Therefore, waveguides that contain isotropic 2D materials to induce plasmonic confinement, are often combined with other 3D layers to enable field confinement in the transverse direction orthogonal to the 2D material utilized. For example, dielectric ridges have been utilized along with such 2D material-enabled plasmonic waveguides \cite{Xu:15,Zheng:2016,Huang:2019}. In these structures, the guided modes are based on oscillating fields within the core, much the same as conventional dielectric waveguides that support guided modes via total internal-reflection. To avoid confusion, here we will use "oscillatory field" to refer to the modal fields supported by 3D dielectric waveguides based on TIR, where the fields are localized around the waveguide center, and "evanescent field" to refer to the modal fields supported by 2D plasmonic waveguides, which typically localize around the waveguide edges.

In contrast, plasmonic waveguides, composed of simply just a 2D guiding layer placed in a uniform background, confines waves using a different guiding mechanism, as the modal fields in the waveguide core are evanescent. We shall call these structures 2D plasmonic waveguides in this work. Because of the evanescent field in the waveguide core, 2D plasmonics realized using isotropic 2D materials are not able to support the full suite of optical functions offered by devices based on 3D materials. One prominent example is in-plane coupling devices that are abundant in 3D dielectric waveguides. Examples include multimode interference (MMI) couplers and arrayed waveguide gratings. Another suite of examples include out-of-plane coupling devices. Grating structures are primarily used in 3D photonics to couple in-plane to out-of-plane radiation \cite{chang:2013,Chen:2014,Carrier:2014}, which suffer from limited bandwidth of operation due to their resonant nature. Currently, an efficient nonresonant coupling structure using 2D materials for in-plane and out-of-plane coupling devices, is still lacking.

Pivoting on recent developments and advances in emerging 2D anisotropic materials \cite{Tony:2014,Lin:2016}, in this work we demonstrate analytically and confirm numerically how to leverage anisotropy to realize 2D plasmonic waveguides with guided modes that utilize fields which are oscillatory in nature inside the core. It is found that by manipulating material absorption and anisotropy, the modal fields of anisotropic 2D ribbon modes can be tailored to resemble those obtained in 3D dielectric waveguides. The ribbon modes propagating with oscillatory behavior therefore can enable efficient interference-based optical components such as multimode interference (MMI) devices, or coupling between planar ribbon modes and slab surface modes all within the same integrated 2D material platform. Furthermore, using anisotropic 2D nanoslots that are structurally complimentary to anisotropic 2D ribbons, we demonstrate that the formation of 2D leaky modes can enable leaky antenna which can effectively couple in-plane propagating modes into radiating modes in a non-resonant manner. Unlike their isotrop ic counterparts, the material absorption of anisotropic 2D films can therefore play a positive role in expanding the capabilities of 2D plasmonic waveguides, enabling the future development and advances of integrated plasmonics, transformation optics, and metasurfaces devices using 2D materials. 

\section{3D Photonic functions in 2D plasmonic materials}
\subsection{Modal fields in anisotropic 2D material ribbon waveguides}
\begin{figure}
	\centering
	\includegraphics[width=0.499\textwidth]{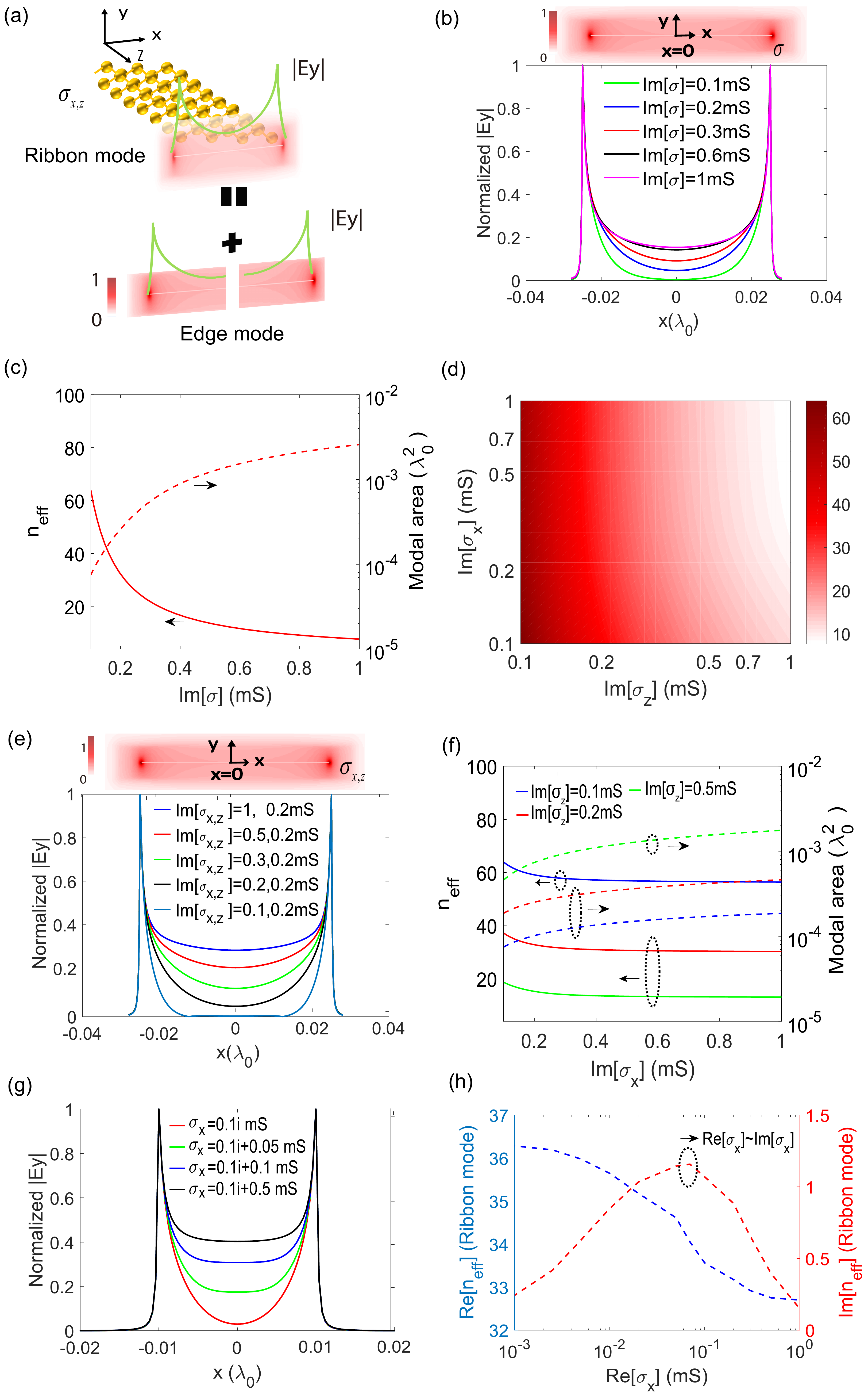}
	\caption{\textbf{(a)} Schematics of the formation of plasmonic ribbon modes supported by 2D materials, which can be considered as a result of the coupling of two edge modes on either side of the film. \textbf{(b,c)} The normalized field distributions, the effective index, and modal area of the isotropic ribbon modes using varying conductivity ($Im[\sigma]$) when the width of the film is $w/\lambda_{0}=0.05$ and $\lambda_{0}=5\mu m$. It is shown that the modal field is pushed toward the centralized film using higher $Im[\sigma]$. Concurrently, increasing the value of $Im[\sigma]$ will lead to smaller $n_{eff}$ and weaker modal confinement. \textbf{(d)} Effective index ($n_{eff}$) of the ribbon modes supported in anisotropic 2D materials schematically shown in (a) as function of in-plane conductivities, which depends more sensitively on the conductivity in the waveguide propagation direction $Im[\sigma_{z}]$. \textbf{(e,f)} The normalized field distributions, the effective index and modal area of the anisotropic ribbon modes using varying degree of in-plane anisotropy. It is seen that the modal field is more contained at the center of the film when $Im[\sigma_{x}]>Im[\sigma_{z}]$, while becomes more localized around the edges when $Im[\sigma_{z}]>Im[\sigma_{x}]$. For anisotropic film, the change in $Im[\sigma_X]$ does not significantly change the $n_{eff}$ and the modal confinement of the ribbon modes. \textbf{(g,h)} The normalized field distributions and effective index of the anisotropic ribbon modes using varying $Re[\sigma_{x}]$. It is seen that ribbon modes can become more oscillatory when $Re[\sigma_{x}]>Im[\sigma_{x}]$. In addition, the modal loss ($n_{i}$) will increase with $Re[\sigma_{x}]$ when $Re[\sigma_{x}]<Im[\sigma_{x}]$, but will start to decrease by further increasing $Re[\sigma_{x}]$ such that $Re[\sigma_{x}]>Im[\sigma_{x}]$. The conductivity of the film is assumed to be $\sigma_{x}=0.1i+Re[\sigma_{x}mS],\sigma_{z}=0.2i mS$ and the width of the film is $w=0.02\lambda_{0}$.\label{TIR}}
\end{figure}
To understand how anisotropy can be utilized to enable ribbon modes of oscillatory fields, it is instructive to initially study the dispersion properties of a z-propagating fundamental ribbon mode supported by a 2D isotropic conducting sheet ($\sigma_x=\sigma_z=\sigma$), which can be considered as the coupling of two edge modes on either side of the film as schematically shown in Fig. \ref{TIR}(a). The plot in Fig. \ref{TIR}(b) depicts the field distributions of the isotropic ribbon modes in the ribbon structure using varying values of conductivity $\sigma$. Here the film is assumed to be free-standing in vacuum and the operational wavelength is $\lambda_{0}$=5$\mu$m, where common 2D materials such as graphene and black phosphorus (BP) are metal-like and can support plasmonic waveguiding. It should be noted that all the results found in this work can also be applicable to other wavelength regimes as long as 2D plasmonics is tenable and to other constructs as long as the 2D films are situated in a uniform medium such as silica as an example. As elucidated in \cite{Chang:20}, using higher value of $\sigma$ will give rise to smaller $n_{eff}$ and leads to slowly decaying edge modes that constitute the ribbon modes. As such, although the modal fields of the isotropic ribbon modes can be slightly pushed toward the center of the ribbon structure when $Im[\sigma]$ is increased from 0.1mS to 1mS, the modal area, defined by $\frac{1}{max(W)}\int WdA, W=\frac{1}{2}Re[\epsilon]\left | E \right |^2+\frac{1}{2}\mu_{0}H^2$, is also seen to increase by two order of magnitudes as a result of smaller $n_{eff}$, as depicted in Fig. \ref{TIR}(c).

In contrast, the field profiles of the 2D ribbon modes can be tailored more effectively through the use of anisotropic ribbon structures (Im[$\sigma_{x}$]$\ne$Im[$\sigma_{z}$]). To illustrate this, the plot in figure \ref{TIR}(e) depicts the field distributions of anisotropic ribbon modes using various values of in-plane conductivities. It is shown that the modal fields contained in the core region can be significantly enhanced when $Im[\sigma_x]>Im[\sigma_z]$. For instance, in the present configuration it is calculated that 50$\%$ of the modal field can be contained in the region where $-0.02<x/\lambda_{0}<0.02$ when $Im[\sigma_x]=1mS$ and $Im[\sigma_z]=0.2mS$, which is higher than those of the isotropic cases when $Im[\sigma_x]=Im[\sigma_z]=0.2mS$ and $1mS$. Note that the values of these anisotropic conductivity tensor can be attained using 20nm BP when the applied chemical potential is 0.2eV. On the other hand, when $Im[\sigma_x]<Im[\sigma_z]$ anisotropic conductivities will result in stronger field localization around the ribbon edges. Clearly, plasmonic ribbon modes of stronger oscillatory nature can be attained through the utilization of anisotropic 2D materials, as long as the value of $Im[\sigma_x]$ is higher than that of $Im[\sigma_z]$ in such materials. As we shall show later, the ability to render oscillatory fields within such ribbon modes can effectively excite slab modes and facilitate in-plane coupling devices in 2D plasmonics.

Notably, the $n_{eff}$ of ribbon modes enabled by $Im[\sigma_x]>Im[\sigma_z$] does not significantly reduce with increasing $\sigma_x$. As shown in Figs.\ref{TIR}(d) and (f), the $n_{eff}$ of anisotropic ribbon modes depends more sensitively on the conductivity along the propagation direction ($Im[\sigma_z]$). Since $n_{eff}$ dictates the modal confinement of the 2D plasmonic ribbon modes, the introduced material anisotropy thus allows one to tailor the field distribution of ribbon modes without negatively impacting the modal confinement. Such a design benefit is not obtainable in isotropic 2D materials, in which the reduction of decay rate of the edge modes into the film can only be achieved with the reducing in $n_{eff}$, which inevitably leads to weaker modal confinement for the ribbon modes as depicted in Fig. \ref{TIR}(c).

It is instructive to note that, the oscillatory ribbon modes can also be formed through the utilization of material absorption of anisotropic 2D materials. As illustrated in Fig. \ref{TIR}(g), by increasing the material absorption in $\sigma_{x}$ ($Re[\sigma]$) the modal fields of anisotropic ribbon modes are seen to be enhanced around the center of the ribbon film. Similar to anisotropic films free of material absorption ($Re[\sigma]=0$), such a field profile can be attributed to the tailoring of the decaying rate of the edge modes enabled by material's anisotropy, but using material absorption. More notably, in the regime where $Re[\sigma_{x}]>Im[\sigma_{x}]$, the modal loss of the anisotropic ribbon modes is seen to decrease with increasing $Re[\sigma]$, thereby allowing for the formation of low loss ribbon modes even when the waveguides are operated in high material absorption regime as depicted in Fig. \ref{TIR}(h).
The material absorption of anisotropic 2D materials, which can be tuned by the applied bias voltage, thus can offer beneficial design capabilities that are not attainable in isotropic counterpart.

The physical mechanisms behind the enhanced coupling of edge modes showcased in Fig. \ref{TIR} can be understood through the field distributions of edge modes that extend within the in-plane 2D film, which can be assessed using \cite{Chang:20}:

\begin{equation}\label{decay}
{q_x} \simeq \sqrt {|\frac{{{\sigma _z}}}{{{\sigma _x}}}|} {n_{eff}},
\end{equation}
where ${q_{x,z}}$ is the wave number in the $x,z$ directions normalized to $k_{0}$ respectively with the waveguide field in the form ${e^{{k_0}\left( {i{q_z}z + {q_x}x} \right)}}$.

For a ribbon waveguide with a fixed waveguide width, $q_x$ dictates the decaying rate of the edge modes that extend into the 2D film. To render oscillatory field within 2D ribbon waveguides, the value of $q_x$ therefore should be reduced to enhance coupling of two edge modes on either side of the film:

(a) In isotropic films ($\sigma_x=\sigma_z=\sigma$), enhanced coupling can only be achieved through the reduction of $n_{eff}$ with the use of smaller value of Im[$\sigma$], which in turn accounts for the evolution of the field distribution profiles observed in Fig. \ref{TIR}(b). However, as shown in Fig. \ref{TIR}(c), the realization of oscillatory field in turn leads to an increase in modal area and therefore compromises the modal confinement the ribbon waveguides

(b)	In anisotropic films, stronger coupling of the two edge modes can be achieved by manipulating the ratio between $\sigma_{x}$ and $\sigma_{z}$ without impacting $n_{eff}$. In particular, as displayed in Fig. \ref{TIR}(e), increasingly oscillatory field distribution can be realized by increasing $Im[\sigma_{x}]$ within anisotropic ribbons. However, contrary to isotropic counterparts, oscillatory field engineering occurs without impacting $n_{eff}$ and modal confinement can be maintained.

(c)	In anisotropic films, modal absorption also provides an additional design freedom for field manipulation. Specifically, as shown in Fig. \ref{TIR}(g) , by increasing the value of $Re[\sigma_{x}]$, it is possible to achieve similar field distributions to those observed in Fig. \ref{TIR}(e). Most importantly, as elucidated in \cite{Chang:20}, increasing $Re[\sigma_x]$ will not lead to an increase in modal loss, because the destructive interference of the edge modes diminishes the Ex field intensity. The ability to leverage material absorption for field engineering without incurring additional modal loss offers additional advantage over isotropic structures.

It is important to highlight that anisotropy has also been utilized to facilitate field and dispersion engineering of 3D metamaterial waveguides. For instance, by incorporating metamaterials, metamaterial waveguide can allow for photonic skin-depth engineering \cite{vanNiekerk:21} and subwavelength mode confinement using transparent materials \cite{Jahani:14}. However, the material absorption inherent in these material constituents is typically regarded as an unwanted bi-product that negatively impacts the mode propagation distance. On the other hand, with anisotropic 2D materials, material absorption instead can have favorable effects on the modal attributes, thus offering a new pathway toward design of waveguide fields unattainable in existing 3D counterparts.

\subsection{In-plane coupling using anisotropic 2D materials}
\begin{figure}
	\centering
	\includegraphics[width=0.499\textwidth]{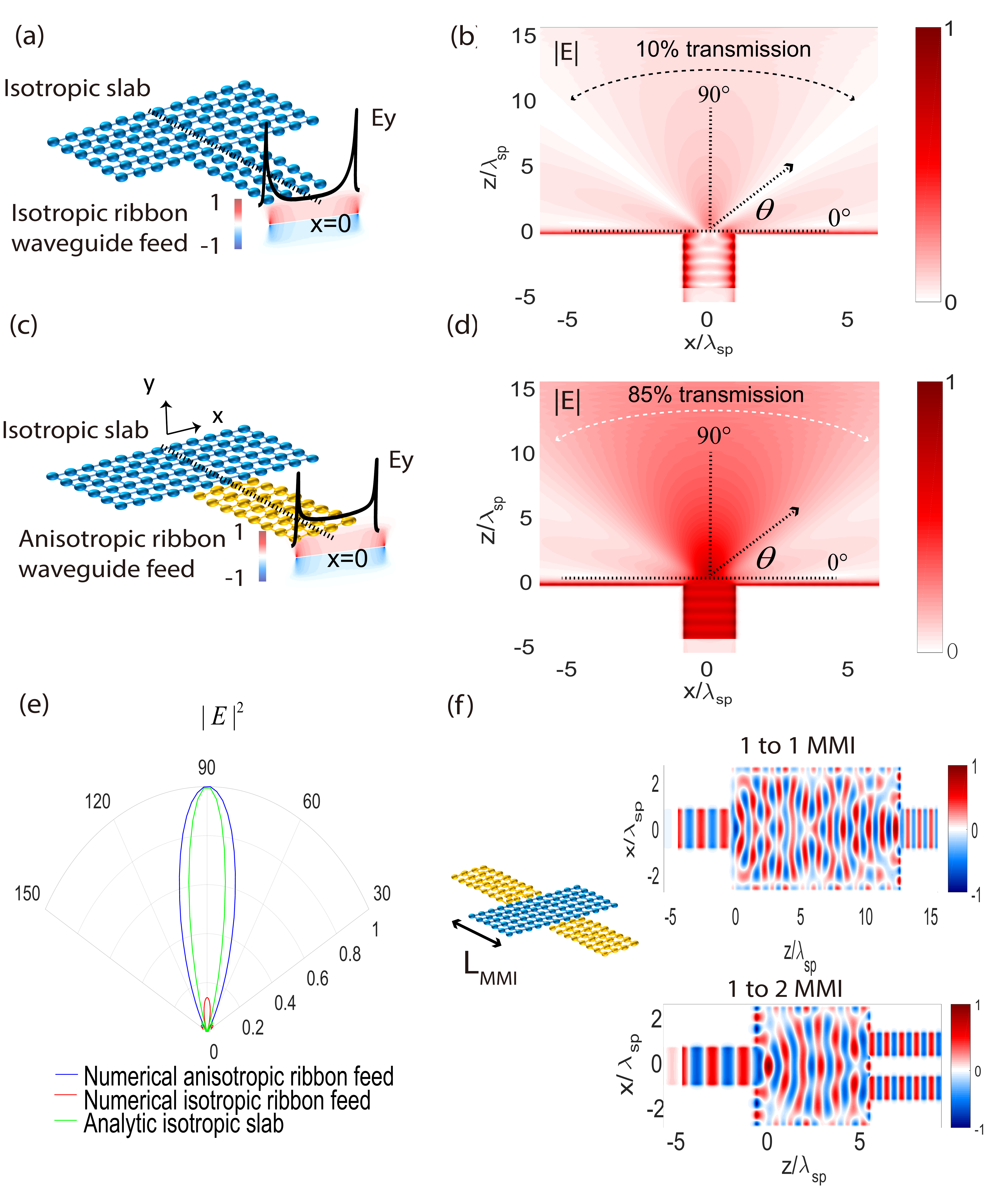}
	\caption{\textbf{(a)} Schematic of an unbounded isotropic 2D material slab connected to a 2D isotropic ribbon waveguide, where the 2D slab mode is fed a isotropic ribbon mode. \textbf{(b)} The electric field profile $|E|$ of an isotropic 2D slab fed by an isotropic ribbon mode. The conductivity of the film is $Im[\sigma]=0.2mS$ and the width of the ribbon is $w/\lambda_{0}=0.08$. Due to the edge mode nature, isotropic ribbon field can only lead to suboptimal coupling efficiency into the slab mode. \textbf{(c)} Schematic of an unbounded isotropic 2D material slab connected to a 2D anisotropic ribbon waveguide, where the 2D slab mode is fed an anisotropic ribbon mode. \textbf{(d)} The electric field profile $|E|$ within an isotropic 2D slab fed by an anisotropic ribbon mode, with conductivities of $Im[\sigma_{x,z}]=1,0.2mS$. It is seen that the ribbon mode of oscillatory field can significantly improve the coupling efficiency to the excitation of the slab mode. \textbf{(e)} Comparison of the radiation patterns in the 2D slab fed by isotropic ($Im[\sigma]=0.2mS$) and anisotropic ribbon modes. The analytical result is also displayed. \textbf{(f)} The schematic and simulated field profiles of the MMI devices using anisotropic 2D materials ($Im[\sigma_{x,z}]=0.5,0.2mS$). Using image theory developed in 3D optics, the propagation distance ($L_{MMI}$) of N-imaging splitters can be approximated by $L_{N}=3L_{\pi}/4N$.}\label{MMI}
\end{figure}

Taking advantage of the sub-wavelength confinement in ribbon modes, concomitantly with modes that propagate possessing oscillatory behavior, anisotropic 2D materials can therefore support optical functions which are abundantly found in optics of 3D materials but can now be realized in 2D plasmonics. One important function discussed here is the ability of the guided field to propagate in a 2D slab within an integrated platform. Recently, 2D material plasmon waves have shown significant promises in wave-mater interaction and manipulations that enable new photonic functionalities in the deep subwavelength regimes \cite{Alu:2016}. As schematically shown in Fig. \ref{MMI}(a), such a platform can be considered when a 2D ribbon waveguide is connected to an unbounded 2D material slab which support 2D plasmonic slab modes. As can be shown in Fig. \ref{MMI}(b), if the 2D slab is fed by an isotropic ribbon mode such as graphene, the majority of the input field will propagate along the edges of the film, which in turn leads to suboptimal excitation of the slab modes.
To effectively render plane-wave-like plasmonic slab mode \cite{Alu:2016,Hossein:2016}, the 2D slab should be fed by an anisotropic ribbon mode with the modal field localized around the waveguide center. For instance, in the present configuration the coupling efficiency of the slab modes fed by a ribbon mode can be significantly increased from 15\% to 85\% when the conductivity tensor is $Im[\sigma_{x,z}]=1,0.2mS$. Additionally, the field pattern in the 2D slab fed by a mode of oscillatory behavior can be considered as the result of the electromagnetic fields excited by an uniform field across the aperture. The analytical expressions of the far field profile into the slab can be given as:

\begin{equation}\label{radi}
E({\theta})=sinc(\pi l\sin(\theta)/\lambda),
\end{equation}
where $\theta$ is the angle between the observation point and the center of the aperture. As plotted in Fig. \ref{MMI}(e), it is seen that Eq. \ref{radi} can show good agreement with the simulated results.

Being able to realize such oscillatory modes, anisotropic 2D materials can also facilitate the design of multimode inteference (MMI) splitter in a 2D plasmonic setting. This can be achieved when the width of the 2D film is restricted to a finite size. For MMI, the input optical modes should effectively excite the fundamental and higher order standing wave modes that form the modal basis for MMI waveguides, as schematically shown in Fig.\ref{MMI}(f). According to the image theory for MMIs in 3D dielectric, when the MMI is excited by a symmetric input field profile, the N-fold images can be formed at the distance $L_{N} = 3L_{\pi}/4N$. where $L_{\pi}=\lambda_{0}/2\delta n$ is the beating distance of the fundamental and the first higher order modes. For proof-of-concept, we study the behavior of several 1-to-1 and 1-to-2 MMIs using 2D anisotropic ribbon structures, which are shown in Fig.\ref{MMI}(f). It is seen that the coupling length of the devices can be deeply in the subwavelength regime thanks to the strong confinement of modes using 2D plasmonics, while utilizing the design knowledge developed in 3D optics. By utilizing  advances in emerging 2D conducting materials, such a device platform can be realized through spatially tailoring the voltage bias on the planar surface \cite{Tony:2014}, or forming the lateral heterostructure with several thin-film 2D materials patterned and etched on the same planar surface \cite{Neto:2016}.
\begin{figure}
	\centering
	\includegraphics[width=0.499\textwidth]{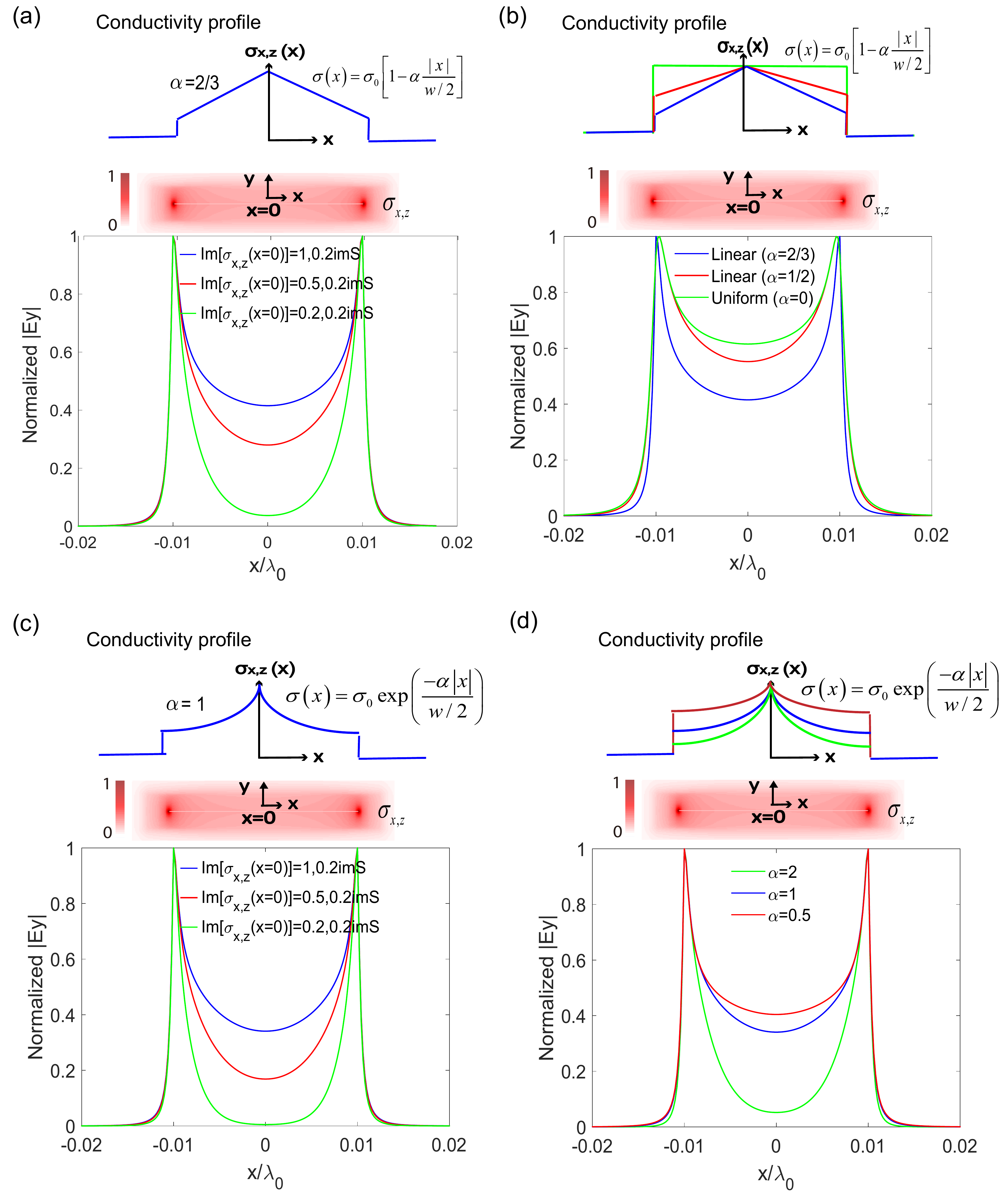}
	\caption{\textbf{(a)} Normalized field distributions of anisotropic 2D ribbon with spatial variation of conductivity tensor using various degree of material anisotropy. The values of conductivity tensor are linearly decreased in the x direction with $\alpha$=2/3. \textbf{(b)} Normalized field distributions of anisotropic 2D ribbon using varying $\alpha$ in the linear conductivity distributions, where $Im[\sigma_{x,z}]=1,0.2imS$. \textbf{(c)} Normalized field distributions of anisotropic 2D ribbon when the values of conductivity tensor are exponentially decreased in the x direction with $\alpha$=1. \textbf{(d)} Normalized field distributions of anisotropic 2D ribbon using varying $\alpha$ in the conductivity distributions, where $Im[\sigma_{x,z}]=1,0.2imS$. In both cases, anisotropic ribbons can still better support oscillatory field distribution as compared to the isotropic counterparts. The linear and exponential reductions in conductivity tensor will lead to the enhancement of edge fields that push the modal field toward the edges of the film.}\label{nonuniform}
\end{figure}
Note that the oscillatory field distribution observed in anisotropic ribbon can also be supported within the ribbon structures with spatially-varying conductivity. For example, Fig. \ref{nonuniform} compares the field distributions of the ribbon modes when the values of the conductivity tensor reduce linearly or exponentially toward the ribbon edges. From Fig. \ref{nonuniform}(a), it is seen that anisotropic ribbons can facilitate oscillatory field despite the spatially varying surface conductivity, which was not observed for isotropic counterpart. Moreover, Fig. \ref{nonuniform}(b) compares the modal field profiles supported by anisotropic 2D ribbon with different linearly varying spatial conductivities. As can be seen, the ribbon structure can better support oscillatory field when the conductivity of ribbon has a uniform distribution, but the linear reduction in conductivity tensor leads to enhancement of edge field due to higher $n_{eff}$. Similar modal behaviors can also be observed when the conductivity of 2D film decays exponentially, as displayed in Figs. \ref{nonuniform}(c) and (d). Overall, the material anisotropy can still facilitate oscillatory field as compared to isotropic counterpart despite surface conductivity being not uniform. Therefore, manipulation of material anisotropy is an practical approach towards ribbon waveguide design and can accommodate unintentional variation in conductivity due to fabrication imperfections.

\begin{figure*}[t!]
	\centering
	\includegraphics[width=0.9\textwidth]{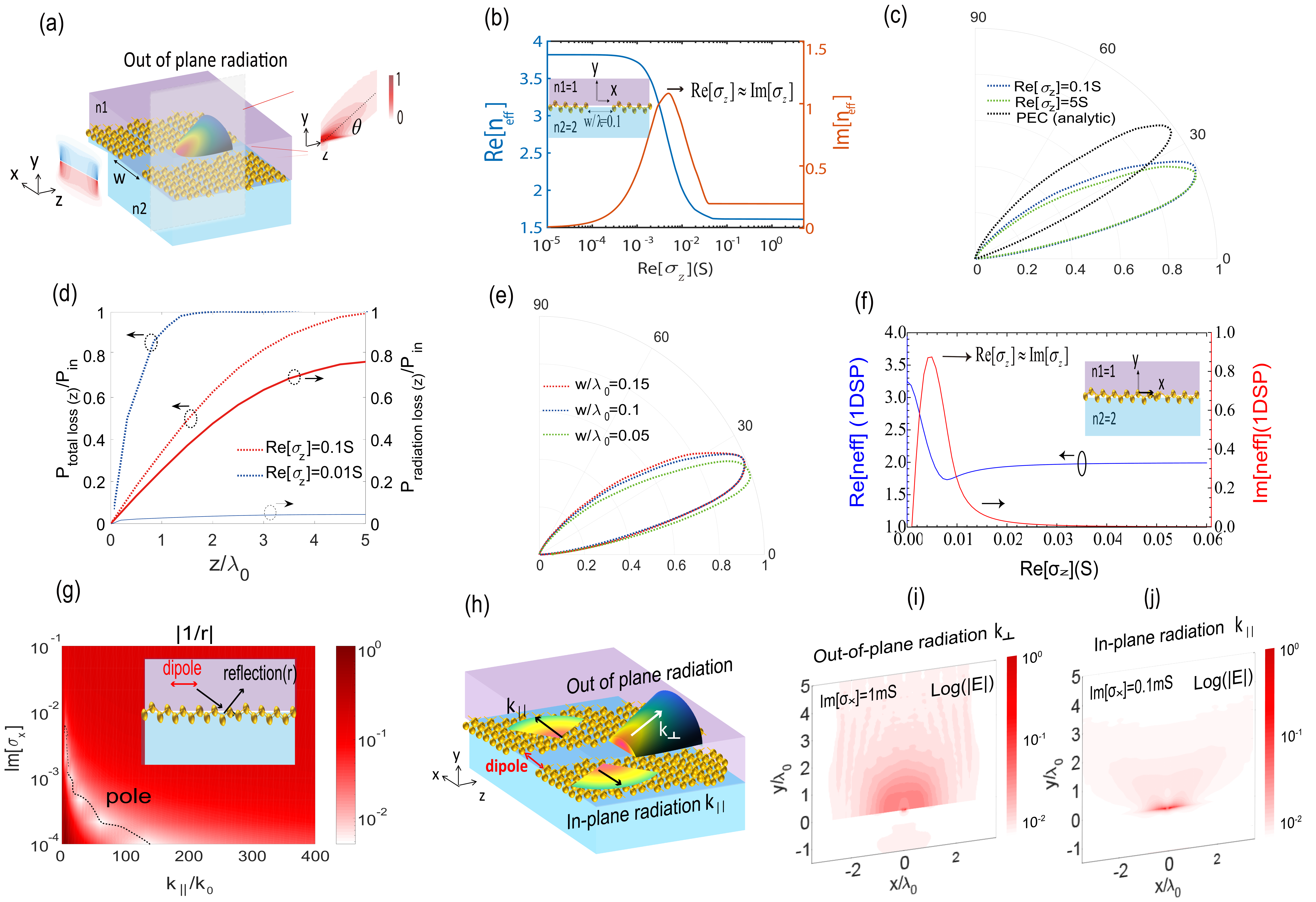}%
	\caption{\textbf{(a)} Schematic of a slot leaky wave antenna utilizing anisotropic 2D materials. The inset shows the electric field radiation pattern in the y-z plane, which radiates into the high-index regime $n_{1}$. The conductivities of the 2D films are $\sigma_{z}=0.005i+\sigma_{zr},\sigma_{x}=0.005i$ and the width of the slot is assumed to be $w/\lambda_{0}=0.1$. \textbf{(b)} Dispersion of the slot mode as function of material absorption ($\sigma_{zr}$). It is seen that the value of $n_{eff}$ of the mode can be smaller than that of the cladding by increasing material absorption. \textbf{(c)} Normalized radiation patterns for the slot leaky wave antenna when $\sigma_{zr}$ of the 2D material is $0.1S$ and $5S$, which shows that the radiation pattern will not change significantly when $Im[\sigma_{z}]>Re[\sigma_{z}]$. \textbf{(d)} Normalized radiation efficiency for the leaky wave antenna when $\sigma_{zr}$=0.01 and 0.1S. It is seen that the material absorption leads to strong propagation loss when $\sigma_{zr}=0.01$ but can result in leaky mode radiation when $\sigma_{zr}=0.01$. \textbf{(e)} Radiation pattern for the slot width ranging from $w/\lambda_{0}=0.05$ to $w/\lambda_{0}=0.15$ when $\sigma_{zr}=0.1S$. The radiation pattern only changes slightly due to the broadband nature of the slot leaky modes. \textbf{(f)} Dispersion of the one-dimensional plasmonic mode supported by an infinite 2D material as function of material absorption ($\sigma_z=0.005i+\sigma_{zr}$). The 2D film is sandwiched between n=1 and n=2 substrate and cladding and the waveguide propagation direction is assumed to be in the z direction, which displays similar dispersion characteristics as those observed in the 2D slot modes. \textbf{(g)} The modulus of the reciprocal of Fresnel coefficient as functions of $k_{\parallel}/k_{0}$ and $Im[\sigma_{x}]$, when the orientation of the dipole source is parallel to the slot. The position of the poles is indicated in the white dashed line. Highly confined surface wave can be excited for smaller $Im[\sigma_x]$. \textbf{(h)} Schematic of the in-plane propagating and out-of-plane radiation fields for leaky wave slot antenna, which can be considered when a dipole source is placed on top of the slot with its orientation being parallel to the slot. \textbf{(i),(j)} The logarithmic scale of the cross sectional field profiles in the x-y plane when $Im[\sigma_x]=1mS$ and $0.1mS$, which depict the field profiles of the radiation leaky mode and the lossy propagating surface mode respectively, with majority of the fields propagating in the out-of-plane and in-plane respectively.}\label{leaky}
\end{figure*}

\subsection{Out-of-plane coupling using anisotropic 2D materials}

In addition to in-plane coupling, another optical function which can be achieved using anisotropy is the radiation of the optical modes supported by 2D plasmonic waveguides into an unbounded plane wave in the out-of-plane direction. Such a function can play a pivotal role in enabling non-resonant, broadband, efficient in-out coupling to-from 2D films, respectively.

To achieve this, we consider a 2D nanoslot structure which is comprised of two semi-infinite anisotropic 2D films separated by a distance $w$. This is structurally complimentary to the 2D ribbon waveguides discussed earlier in this work. The optically narrow 2D slot is sandwiched between a high index upper cladding ($n_1$) and a low index substrate ($n_{2}$), as schematically illustrated in Fig. \ref{leaky}(a). To obtain optimal radiation efficiency, the 2D films here need to be operated in the high material absorption regime \cite{Neto:2003,Wang:11}. The high degree of the tunability in 2D material's conductivity thus can make 2D materials particularly well suited for this function.

The dispersion of the 2D plasmonic slot mode is investigated as function of the material absorption in the propagation direction, which is depicted in Fig. \ref{leaky}(b). The conductivity of the film is assumed to be $\sigma_{z} =0.005i+\sigma_{zr}$; $\sigma_{x} =0.005i$ and the width of the slot is $w/\lambda_{0}=0.1$, where $\sigma_{zr}$ denotes the real part of $\sigma_{z}$. It can be seen that the propagation loss of the guided 2D slot will increase as $\sigma_{zr}$ increases in value. However, by further increasing material absorption such that $\sigma_{zr}>\sigma_{zi}$, the slot mode can become a leaky mode. In this case, the $Re[n_{eff}]$ of the slot mode is reduced below the high-index cladding but higher than that of low-index substrate. For even higher $\sigma_{zr}$ ($\sigma_{zr}>>\sigma_{zi}$), the values of $n_{eff}$ become asymptomatic closer to those obtained from PEC.
\begin{figure}[t!]
    \centering
    \includegraphics[width=0.49\textwidth]{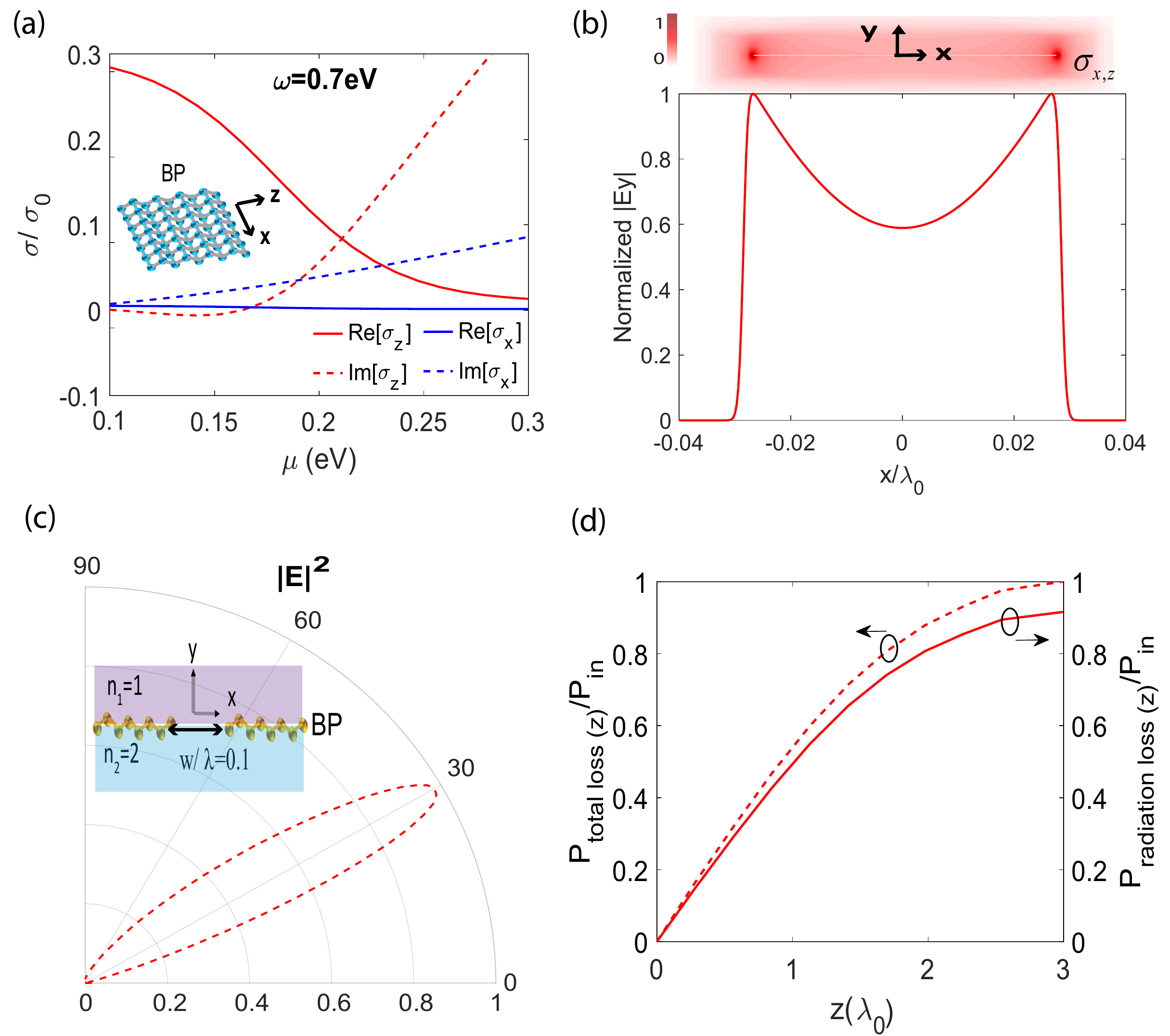}
    \caption{\textbf{(a)} Conductivity of 20nm BP in the armchair direction ($\sigma_{z}$) as function of chemical potential, operated at $\omega=0.7eV$. It is seen that the value of $Re[\sigma_z]$ can exceed that of $Im[\sigma_z]$ when $V<0.22eV$. \textbf{(b)} Field profile of BP ribbon structure when BP’s crystal axes in the zigzag direction is to align with the direction of light propagation, which can render oscillatory field distribution discussed in this work. \textbf{(c)} Radiation pattern and \textbf{(d)} radiation efficiency of the BP-based leaky wave slot antenna when BP's armchair direction is alined with the lightwave propagation direction. The width of the slot is $w/\lambda_{0}=0.1$. The functional device can be highly directive and can attain radiation efficiency of $90\%$} \label{leaky2}
\end{figure}

When utilized in the proposed slot leaky wave design, the 2D materials should be operated in the $\sigma_{zr}>\sigma_{zi}$ region. The plot in Fig. \ref{leaky}(c) shows the radiation pattern of the slot leaky wave antenna using $\sigma_{zr}=0.1S$ and $5S$. The radiation pattern based on PEC is also plotted for comparison. It can be seen that the slot leaky wave antenna using 2D materials can be directive with 3dB beamwidth of $28$ degree. It is noteworthy to point out that the material absorption can impart different effects onto the slot mode depending on the values of $\sigma_{zr}$. As shown in Fig. \ref{leaky}(d), when $\sigma_{zr}=0.01S$ the material absorption in this case will lead to strong propagation loss with only minimal field radiated. However, when $\sigma_{zr}$ is increased to $0.1S$, material absorption can lead to significant radiation loss, where the majority of the leaky mode will be radiated into the far field instead of being dissipated.

One prominent feature associated with the leaky wave slot antenna is its broadband nature \cite{Wang:11}. It can be observed in Fig. \ref{leaky}(e) that the radiation pattern only changes slightly with the slot width. As such the device is operable over a broad bandwidth and more tolerant to fabrication errors. Utilizing material absorption of 2D materials can therefore facilitate the design of non-resonant optical antennas to efficiently couple guided modes in 2D materials out-of-plane for 3D integration.

To understand how material absorption can lend support to leaky modes, we study the dispersion behavior of a one-dimensional plasmonic mode supported by an infinite-extended 2D material as schematically shown in the inset of Fig. \ref{leaky}(f). Using the transfer-matrix method (TMM), the dispersion of such a one-dimensional plasmonic waveguide can be obtained:
\begin{equation}
\frac{{{\varepsilon _2}\sqrt {n_{eff}^2 - {\varepsilon _1}} }}{{{\varepsilon _1}\sqrt {n_{eff}^2 - {\varepsilon _2}} }} = \frac{1}{{\sqrt {\left( {n_{eff}^2 - {\varepsilon _2}} \right)} \frac{{{\eta _0}\sigma_{z} }}{{{\varepsilon _2}}} - 1}},\label{TMM}
\end{equation}
where $\varepsilon_{1,2}$ denote the refractive indices of the cladding and substrate respectively, and $\sigma_{z}$ is the conductivity of anisotropic 2D films along the waveguide propagation direction.

The plot in Fig. \ref{leaky}(f) examines the dispersion characteristics of a one-dimensional plasmonic mode supported by 2D materials as function of material absorption ($Re[\sigma_z]$), where the imaginary part of $\sigma_{z}$ ($\sigma_{zi}$) is assumed to be $0.005S$ and $n_{1,2}=2,1$. It is seen that in the region where $\sigma_{zr}<\sigma_{zi}$, the real part of $n_{eff}$ will decrease while the imaginary of $n_{eff}$ will increase with increasing $\sigma_{zr}$. On the other hand, $Re[n_{eff}]$ will increase while $Im[n_{eff}]$ will decrease by further increasing $\sigma_{xr}$ such that $\sigma_{zr}>\sigma_{zi}$. From Eq. \ref{TMM}, it is shown that to form the leaky modes, the material absorption should be introduced in the conductivity tensor along the lightwave propagation direction, which in turn give rise to the 2D slot leaky modes demonstrated in Fig. \ref{leaky}.

Note that for a 2D film without material absorption ($Re[\sigma]=0$), $n_{eff}$ calculated using Eq. \ref{TMM} is always larger than those of the cladding and substrate. As a result, the 2D film will serve as a waveguiding structure and support guided mode that does not radiate. To support leaky modes, the condition $n_{1}<n_{eff}<n_{2}$ needs to be satisfied in order for optical modes to radiate into the dense medium ($n_2$). This is possible by regulating the material absorption of anisotropic 2D materials such that $Re[\sigma_z]>Im[\sigma_z]$ [Fig. \ref{leaky}(f)]. For $Re[\sigma_{z}]>>Im[\sigma_{z}]$, $n_{eff}$ can be approximated by the average between $n_1$ and $n_2$ since the anisotropic 2D material is operated near the PEC regime \cite{Neto:2003}.

Notably, the conductivity in the lateral direction ($\sigma_{x}$) may also affect the radiation characteristics. To understand this effect, it is instructive to consider the radiation pattern caused by a dipole placed on top of the slot \cite{Neto:2003} when its orientation is parallel to the slot, as schematically shown in Fig.\ref{leaky}(h). For finite conductivity $\sigma_{x,z}$, the emitted dipole therefore will not only give rise to the radiated field that propagates in the out-of-plane direction ($k_{\bot}$), but also the in-plane bounded surface waves ($k_{\parallel}$) propagating along the 2D materials. The Fresnel reflection coefficient of a dipole source impinging from medium $n_{1}$ to medium $n_{2}$ can be derived as \cite{Zhan:2013}:

\begin{equation}
r = \frac{{{n_1}/{q_1} - {n_2}/{q_2} - \overline {{\sigma _x}} }}{{{n_1}/{q_1} + {n_2}/{q_2} + \overline {{\sigma _x}} }},\label{Refl}
\end{equation}
where ${q_{1,2}} = \sqrt {k_{1,2}^2 - k_\parallel ^2}$ and $\overline\sigma_{x}=\sigma_{x}\eta_{0}/k_{0}$ is the normalized conductivity transverse to the waveguide propagation direction.

The wave vector of the in-plane surface waves then can be obtained by finding the poles of the Fresnel equation as given by Eq. \ref{Refl}, which yields:
\begin{equation}
\frac{{{\varepsilon _2}}}{{\sqrt {k_1^2 - k_\parallel ^2} }} + \frac{{{\varepsilon _1}}}{{\sqrt {k_2^2 - k_\parallel ^2} }} + \frac{{{\sigma_x}{\eta _0}}}{{{k_0}}} = 0.\label{pole}
\end{equation}

The plot in Fig.\ref{leaky}(g) shows the modulus of the reciprocal of Fresnel reflection coefficient as functions of $Im[\sigma_{x}]$ and in-plane wave vector ($k_{\parallel}/k_{0}$). It is seen that the position of the poles in Eq. \ref{Refl} will shift to higher spatial harmonics ($k_{\parallel}$) with smaller $Im[\sigma_{x}]$, which in turn can lead to the excitation of in-plane surface wave.

\subsection{2D plasmonic devices using Black phosphorus}

By dynamically tuning the complex anisotropic conductivity of the 2D films, it is possible to tune the dispersion of the gap modes. This effect can be gate tunable as the conductivity can be tuned using the potential applied with the aim of switching on or off the radiation function. Such ability cannot be achieved in previous work that utilized isotropic media to statically tuned the dimensions of the structure to regulate the real propagation constant of such modes to increase their value above that of the superstrate or substrate in order to induce radiation.

Specifically, the tuning of $Im[\sigma_{x}]$ can be chosen to suppress the coupling of the in-plane surface waves. The plots in Figs.\ref{leaky}(i) and (j) illustrate the radiation profile of the proposed leaky wave slot antenna using $Im[\sigma_{x}]=1mS$ and $0.1mS$ when $\sigma_{z}=0.01+0.05i$ and $w/\lambda_{0}=0.1$. It can be observed that when $Im[\sigma_{x}]=0.1mS$ the majority of the modal field will propagate in the form of the surface plasmon wave, whereas in the case of $Im[\sigma_{x}]=1mS$ most of the optical field will radiate in the out-of-plane direction.

To place our results into perspective, natural anisotropic 2D materials such as black phosphorus (BP) \cite{Tony:2014,Lin:2016} can be considered for the implementation of the slot leaky wave antenna shown in this work. For instance, the plot in Fig. \ref{leaky2}(a) depicts the conductivity of 20nm BP in the armchair direction ($\sigma_{z}$) as function of the applied chemical potential at $\omega=0.7eV$. It is seen that the value of $Re[\sigma_z]$ can become larger than that of $Im[\sigma_{z}]$ when the chemical potential is below 0.22eV. For BP, it is seen that the material absorption of BP in the armchair direction can be tuned more significantly compared to the zigzag direction. Moreover, BP can operate in the low ($Re[\sigma_{z}]<Im[\sigma_{z}]$) or high material absorption ($Re[\sigma_{z}]>Im[\sigma_{z}]$) regimes depending on bias. If BP’s zigzag axis is aligned with the direction of light propagation, the oscillatory field distribution can be attained [Fig. \ref{leaky2}(b)]. On the other hand, with the armchair direction aligned to the direction of light propagation, leaky modes can be formed using a BP slot waveguide that is structurally complementary to BP ribbon (based on Eq. \ref{TMM}). Therefore, BP is suitable for implementing the various photonic coupling functions discussed in this work. Note that the strong tunability of BP’s absorption in the armchair direction ($Re[\sigma_{z}]$) has also been utilized for optical modulation \cite{Lin:2016}. Figures \ref{leaky2}(b) and (c) show the radiation characteristics of such a BP-based plasmonic slot waveguide operating at $\omega=0.7eV$ and $V=0.1eV$. It is seen that the functional antenna can be highly directive and can achieve radiation efficiency of $90\%$. The high material absorption of anisotropic 2D materials therefore can facilitate design of leaky wave modes in 2D plasmonics with no need to relying on complimenting 3D optics to achieve the desired functions.

In addition to material absorption, the degree of anisotropy of BP’s conductivity can also be tuned significantly, allowing BP to be operated in isotropic or highly anisotropic regimes through electrical bias. Therefore, the applications shown in Fig. \ref{MMI} can also be considered using BP ribbons, which further highlights the wide-applicability of BP ribbons in achieving various integrated photonic functions by leveraging material anisotropy. Table 1 summarizes the values of conductivities used in this work which can be considered using BP. The production of BP nanoribbons can be considered using electron-beam lithography, which can effectively produce phosphorene ribbons with minimal width of 60nm \cite{Ang:2018}. Recently, high quality, individual BP ribbon can also be successfully produced using ionic scissoring of BP crystals, enabling the production of single-atom BP ribbon with width as small as 4nm on different substrates \cite{Watts:2019}.
\begin{table}
\center
\begin{tabular}{ |p{2.7cm}||p{0.7cm}|p{0.7cm}|p{2cm}| }
\hline
Platforms & $\sigma_x$(mS) & $\sigma_z$(mS) & Potential material\\
\hline
Anisotropic ribbon [Figs. \ref{TIR}, \ref{MMI}]  & 1i & 0.2i & 20nm BP (0.3eV)\\
Anisotropic ribbon [Figs. \ref{TIR},\ref{MMI}] &  0.5i& 0.2i & 20nm BP (0.2eV)\\
Anisotropic ribbon/slot [Fig. \ref{leaky2}] & 0.28 & 0.001i & 20nm BP (0.1eV)\\
\hline
\end{tabular}
\caption{The conductivities of the anisotropic 2D material structures used in this work, which can be considered using naturally occurring anisotropic 2D materials such as BP.}
\end{table}

\section{Conclusion}
In conclusion, we have elucidated the potential of anisotropic 2D materials as an emerging class of 2D material platform for integrated plasmonics, which provides an effective tool to tailor the modal fields in 2D material waveguides while simultaneously maintaining deeply subwavelength scale confinement. It is found that oscillatory waveguide modes that are similar to 3D dielectrics can be attained using anisotropic 2D ribbon structure with the aid of material's anisotropy and material absorption. On the other hand, anisotropic 2D nanoslot waveguides can be deployed to support nonresonant 2D leaky modes, which is operable over broad bandwidth and allows for efficient out of plane radiation. Anisotropic 2D materials such as BP therefore can be configured to facilitate in-plane and out-of-plane couplings in 2D plasmonics. The afforded coupling capabilities thus unlock beneficial design advantages in 2D plasmonics that are not attainable in their isotropic counterparts. Furthermore, the use of such reconfigurable materials can also offer a design paradigm where one can dynamically tailor the modal fields of 2D modes in real time, enabling the development of tunable guided modes for active photonic functions, metasurfaces, and transformation optics in 2D materials.



\section*{Acknowledgment}
The authors would like to thank Prof. Nader Engheta from the University of Pennsylvania for the fruitful comments and discussions on this work.

\ifCLASSOPTIONcaptionsoff
  \newpage
\fi



%
\bibliographystyle{IEEEtran}

%







\end{document}